# Albert Einstein at the Zürich Polytechnic: a rare mastery of Maxwell's electromagnetic theory

Galina Weinstein

Einstein at the Zürich Polytechnic: he skipped classes, did not attend all the lectures of his Professors, and before going to the examinations he studied instead from the notebooks of his good friend from class, Marcel Grossmann. Einstein the free-thinker did not respect the two major professors in the Polytechnic – Heinrich Friedrich Weber and Jean Pernet – who eventually turned on him. He felt that his beloved science had lost its appeal because Weber's lectures did not include Maxwell's electromagnetic theory. Einstein seldom showed up to Pernet's practical physics course. By his fourth-rightness and his distrust of authority he had alienated his professors, especially Weber, who apparently conceived a particular dislike of him. At the Zürich Polytechnic, Einstein could not easily bring himself to study what did not interest him. Most of his time he spent on his own studying Maxwell's theory and learning at first hand the works of great pioneers in science and philosophy: Boltzmann, Helmholtz, Kirchhoff, Hertz, Mach. Eventually, Einstein finished first in his class in the intermediate exams; second after him was his note taker Grossmann. Do not ever try Einstein's recipe for studying in college, because after obtaining the diploma, when he sought university positions, he was rebuffed. Finally rescue came from Grossmann, and thanks to him and his father Einstein obtained a post in the Patent Office. There are strong reasons to believe that it was Einstein's rare mastery of Maxwell's electromagnetic theory that ultimately prompted the Director of the Patent Office to offer him a job.

## 1. Secondary school in Aarau

At the beginning of October 1895, at the age of 16½, Einstein went to the Polytechnic institute in Zürich to pass the entrance examination[1]. Einstein lacked a certificate from the Munich Gymnasium (*Maturitütszeugnis*) and was two years under the regular admission age of 18. With the aid of Gustav Maier, a family friend, he received permission from the Polytechnic Director Albin Herzog to take the entrance examination required of applications without certificate.[2]

---

[1] Reiser, Anton, *Albert Einstein: A Biographical Portrait*, 1930/1952, New York : Dover, 1930, p. 44; Winteler-Einstein Maja, *Albert Einstein –Beitrag für sein Lebensbild*, 1924, reprinted in abridged form in *The Collected Papers of Albert Einstein  Vol. 1: The Early Years, 1879–1902* (*CPAE*, Vol. 1), Stachel, John, Cassidy, David C., and Schulmann, Robert (eds.), Princeton: Princeton University Press, 1987, pp xlviii-lxvi; p. xxii; *CPAE*, Vol 1, p. lxv.
[2] Albin Herzog to Gustav Maier, September 25, 1895, *CPAE*, Vol. 1, Doc. 7.

Einstein was seeking to enroll in the engineering section of the Polytechnic. The examination began on October 8 1895 and extended over several days. The results were announced on October 14, 1895. Einstein failed to gain admission. He did so well at his autodidactic preparations, that he passed the entrance examination with the best outcome in mathematical and scientific subjects, but obtained inadequate results in linguistic and historical ones.[3]

Einstein's parents were advised to have their son attend the final year of a Swiss secondary school. His parents were told to enroll Albert to this secondary school, with the prospect of a certain admission the following year, despite the fact that Albert would still be fully six months below the prescribed age (18 years).[4]

Einstein opens his *Autobiographical Skizze* of 1955 (written a month before he died) by recounting the examinations he took on October 1895. He then says that the physicist Heinrich Friedrich Weber invited him to attend his college physics lectures, provided he remained in Zürich. However, continues Einstein, he enrolled according to the rector's advice, Professor Albin Herzog, to the Cantonal School – Kantonsschule – in Aarau, a small Swiss town near Zürich whose schools had a high reputation and as a result were often attended by foreigners, even by some from overseas.[5]

From the end of October 1895 to the early autumn of 1896 Einstein was a pupil in the third and fourth classes in the technical department of the Aargau Kantonsschule. The Kantonschule was originally founded as a private school in 1802 and was taken over in 1813 by the "state" – the Kanton Aargau.[6]

The Aargau Kantonsschule was a "Realschule" (specializing at sciences), not a Gymnasium (specializing at linguistic and humanistic studies).[7] Einstein thus could study the topics he really loved. In addition, he went to a democratic school. The combination of these two traits would lead Einstein to feel at home in the new school, "The system of teaching was liberal, unburdened by too much authority, and resembled university lectures more than high school institution. Each class was not confined to one room, but there was a room for each subject, so that the students moved about for the different subjects as at the university, divided according to subject at definite hours as in college. Furthermore, the teachers were enlightened, modern human beings, Albert immediately felt at home in this circle, made friends with his sound and happy school-fellows and proceeded to enjoy Aarau […]".[8]

---

[3] Winteler-Einstein, *CPAE*, vol 1, 1924, p. xxii; *CPAE*, vol 1, p. 1xv; *CPAE*, Vol. 1, "ETH Entrance Examination and Aargau Kantonsschule", pp. 10-11.
[4] Winteler-Einstein, *CPAE*, Vol. 1, 1924, p. xxii, p. 1xv.
[5] Einstein, Albert, "Erinnerungen-Souvenirs", *Schweizerische Hochschulzeitung* 28 (Sonderheft), 1955, pp. 145-148, pp. 151-153; Reprinted as, "Autobiographische Skizze" in Seelig Carl, *Helle Zeit – Dunkle Zeit. In memoriam Albert Einstein*, 1956, Zürich: Branschweig: Friedr. Vieweg Sohn/Europa, pp. 9-17; p. 9; Winteler-Einstein, 1924, *CPAE*, Vol. 1, p. xxii, p. 1xv.
[6] Correction to Seelig's biography, item 39 084, Einstein's Archives.
[7] Seelig Carl, *Albert Einstein: A documentary biography*, Translated to English by Mervyn Savill 1956, London: Staples Press, p. 12; Seelig Carl, *Albert Einstein; eine dokumentarische Biographie*, 1954, Zürich: Europa Verlag, p. 14.

In the 1955 Skizze Einstein recalled that the "liberal spirit" of Aargau Kantonsschule had imposed an "unforgettable impression" on him. "By comparison with the six years training at the Deutsche, authoritative Gymnasium, I have become aware", says Einstein that "real democracy is not an empty delusion".[9]

In the democratic and free-thinking environment of the Argau Kantonsschule, Einstein imagined a "kindliche Gedanken-Experiment, das mit der speziellen Relativitätstheorie zu tun hat" ("childish thought experiment that has to do with the special theory of relativity").[10] This was the chasing after light beam thought experiment.

Einstein graduated the Aarau secondary school. On November 7, 1896 Einstein sent his Curriculum vita to the Argau authorities:

"I was born on 14 March 1879 in Ulm and, when one year old, came to Munich, where I remained till the winter of 1894-95. There I attended the elementary school and the Luitpold secondary school up to (but not including) class 7. Then, till the autumn of last year, I lived in Milan, where I continued my studies on my own. Since last autumn I have attended the Cantonal school in Aarau, and I now take the liberty of presenting myself for the graduation examination. I then plan to study mathematics and physics in division 6 of the Federal Polytechnic Institute".[11]

**2. Polytechnic in Zürich**

**2.1 Profs. Weber and Pernet turn on Einstein**

In October 1896 Einstein enrolled in division VI of "School for Specialized Teachers in the Natural Science" of the Polytechnic in Zürich.[12] He was almost eighteen and was one of the youngest students to have entered the Polytechnic.

The institute had two chairs, one for Mathematical and Theoretical Physical, held by Professor Heinrich Friedrich Weber, and the other for Experimental Physics, held by Professor Jean Pernet – Eventually both professors Weber and Pernet turned on Einstein. Einstein registered to all of Weber's lectures and laboratory courses. He wrote to Mileva Marić in February 1898, "Weber lectured masterfully on heat (temperature, heat qualities, thermal motion, dynamic theory of gases). I eagerly anticipate every class of his".[13] Einstein nevertheless thought that Weber's "lectures were indeed a little old-fashioned".[14]

---
[8] Reiser, 1930, p. 46.
[9] Einstein, 1955, pp. 9-10.
[10] Einstein, 1955, p. 10.
[11] Dukas Helen, and Hoffmann, Banesh, Albert Einstein, *The Human Side, New Glimpses from his Archives*, Princeton: Princeton university Press.1979, pp. 8-9; p. 121.
[12] "Schule für Fachlehrer mathematischer und naturwissenschaftlicher Richtung".

Weber was a professor of mathematical and technical physics at the Zürich Polytechnic, and he retreated from scientific research. His main enthusiasm was to construct, equip and direct a new physics institute for the polytechnic in Zürich. From the narrow professional point of view being the director of the physics institute at the Zürich Polytechnic was the outstanding event of Weber's career.[15]

Between 1871 and 1874 Weber worked as assistant of Hermann von Helmholtz in the University of Berlin. Weber helped Helmholtz set up and equip the Berlin laboratory and also helped him direct the student laboratories. Helmholtz's laboratory was the first substantial and major laboratory in which Weber worked. It gave Weber the means to conduct both laboratory instruction and research. Weber explicitly acknowledged Helmholtz as his teacher and scientific pathfinder. Indeed, Helmholtz had been one of the leading German scientists. Helmholtz also worked with Albert Michelson.

Weber's work was essentially empirical; he used to develop empirical laws, and empirical theory to explain empirical findings. In addition he was not a man of pure science; even when he developed some theory, he always sought to turn his results to practical and technical use for sources of illumination, telephony, and the establishment of electrical standards.[16] Weber was in his forties when Einstein the fond of pure science entered the polytechnic.

As a regular student in the mathematical and natural science school for teachers, from the winter term 1896-1897 to the summer term 1900, Einstein registered to Weber's courses. Einstein took Weber's following courses and laboratories in the Polytechnic: Physics, Principles and Methods of Measurements in Electrotechnology, Oscillations, Electrotechnical Laboratory, Scientific Work in the Physics Laboratories, Introduction to Electromechanics, Alternating Currents, Ac Systems and Dc Motors, System of Absolute Electrical Measurements, and Introduction to the Theory of Alternating Current.[17]

In 1888 Weber essentially ceased doing scientific research. His publications thereafter were limited to technical reports. Between 1883 and 1902 he published six articles on highly practical, technological subjects, including studies of the transmission of electrical energy between various cities, reports on the use of alternating current systems in electrical railroads, and a report on Swiss federal law concerning low-and high voltage equipment. During Einstein's years at the Polytechnic, Weber published only one scientific paper, and that on a topic closely related to electro-technological concerns: alternating currents.[18]

---

[13] Einstein to Marić, February 16, 1898, *CPAE*, Vol. 1, Doc. 39; Renn, Jürgen. and Schulmann, Robert, *Albert Einstein Mileva Marić The love letters*, translated by Shawn Smith, 1992, Princeton: Princeton University Press, letter 2 .
[14] Reiser, 1930, p. 48.
[15] Cahan, David, "The Young Einstein's physics Education: H.F. Weber, Herman von Helmholz, and the Zurich Polytechnic Physics Institute", in Howard, Don and Stachel, John (eds.), *Einstein the Formative Years, 1879 – 1909: Einstein Studies, Volume 8*, 2000, New York: Birkhauser, pp. 43-82, p. 51.
[16] Cahan, 2000, in Howard and Stachel, pp. 45-50.
[17] *CPAE*, Vol. 1, Doc. 28.

Weber like Helmholtz also was busily in building laboratories. He thus entered into intellectual stasis. By 1890 he had apparently lost interest in and contact with the forefront of research in physics, at the very least in the field of electromagnetism. His lectures of electromagnetic phenomena and laws of electromagnetism remained that of a previous generation. It seems that had he wanted to occupy himself with some empirical research, he did not have much time left to keep pace with modern research, the latest state of science; because from the late 1890's, his preoccupation with building his new institute and with his practical or technical studies apparently left him insufficient time and energy to become adequately acquainted with innovations in science;[19] especially with Maxwell's results and their implications for the foundations of physics and practical physical problems. This was the state of affairs during Einstein's years at the Polytechnic.[20]

Hertz demonstrated electric waves and thus Maxwell's theory began to gain recognition, even though in German speaking countries Maxwellian ideas were accepted more slowly. Change was not immediate and abrupt, but gradually Europe's leading physicists discussed Maxwellian ideas, and especially Helmholtz (Weber's mentor) was the first to embrace Maxwell's fundamental ideas. This was the state of affairs in the late 1880s.

Professor Adolf Fisch who had been in the parallel class at the Argau Kantonsschule with Einstein told Einstein's biographer, Carl Seelig, about the position of physics in those days in the Zürich Polytechnic: "Physics was not in particular favor. Experimental physics was insignificant. Theoretical physics was taught by the electrotechnics professor, Weber. His lectures were outstanding and a magnificent introduction to theoretical physics but Weber himself was a typical representative of classical physics. Anything that came after Helmholtz was simply ignored. At the close of our studies we knew all the past of physics but nothing of their present and future. We were recommended to study the newer literature in private".[21]

Dr. Joseph Sauter, Weber's assistant and later Einstein's colleague at the Bern Patent Office, recalled that "this theory [Maxwell's] was not yet on the official program of the Zürich Polytechnic School".[22]

---

[18] Cahan, 2000, in Howard and Stachel, p. 51.
[19] Cahan, 2000, in Howard and Stachel, pp. 51-52; p. 54.
[20] Cahan, 2000, in Howard and Stachel, p. 52.
[21] Seelig, 1956, p. 29; Seelig, 1954, p. 34.
[22] Sauter, Joseph, "Comment j'ai appris à connaître Einstein", 1960, in Flückiger, Max (1960/1974), *Albert Einstein in Bern*, 1974, (Switzerland: Verlag Paul Haupt Bern), p. 154.

Einstein the free-thinker did not respect Weber, who seemed to have been quite the opposite. Banesh Hoffmann and Helen Dukas say, Einstein's "beloved science had lost its appeal. By his fourth-rightness and his distrust of authority he had alienated his professors, among them Heinrich Friedrich Weber, who apparently conceived a particular dislike of him. This was the same Heinrich Weber who, five years before, had generously gone out of his way to encourage the youth who had failed the entrance examinations. The relationship had since deteriorated, Weber on one occasion saying to Einstein with probably justified exasperation, "You're a clever fellow! But you have one fault. You won't let anyone tell you a thing. You won't let anyone tell you a thing".[23]

Seelig says, "I have learned from many quarters that this teacher had a particular dislike for Einstein. The reason for this may have been that Einstein persisted in calling him 'Herr Weber' instead of 'Herr Professor'. Such trifles can often give rise to instinctive dislike."[24]

It appears that Einstein did not get along with another professor in the Poly, the French Swiss experimental physics Professor Jean Pernet, which he took his practical laboratory courses. Joseph Sauter told Seelig that as a student Einstein worked in the laboratory of Professor Jean Pernet. One day in June 1899 Einstein worked in Jean Pernet's laboratory and he seriously damaged his right hand in an explosion.[25]

Sauter told Seelig the following story. In common with every other experimenter Einstein had a chit on which was written his task and the method to be employed. With his usual independence Einstein naturally flung the paper in the waste paper basket and started to solve the problem in some other than the official way. Furious at this, Pernet once asked his assistant, Schaufelberger, "What do you make of Einstein? He always does something different from what I have ordered". The assistant replied: "He does indeed, Herr Professor, but his solutions are right and the methods he uses are always of great interest."[26]

Einstein seldom showed up to Pernet's practical physics course. As a result in his course "Introduction to the practice of physics – elementary practice of physics", Pernet gave him the lowest possible grade, 1, and entered the only sanction in his Polytechnic record: "March 1899: reprimand from the Director on account of lack of diligence in the Physics Practicum".[27]

Margarette von Uexküll was a fellow student of Einstein, studying biology and she and Einstein's Serbian classmate and future wife Mileva Marić, shared the same lodgings at the house of Johanna Bächtold.[28] Thirty years after Einstein's studies at the Poly, von Uexküll reported that Einstein told her that Professor Pernet had once kindly given him food for thought by saying: "There is no lack of eagerness and goodwill in your work, but a lack of capability". Pernet told Einstein he had no idea how difficult was the path of physics; hence: "Why don't you study medicine, law or philology instead?" "Because I feel that I have a talent, Herr professor" replied Einstein. "Why shouldn't I least try with physics?" "You can do what you like, young man" said Pernet abruptly closing the conversation. "I only wanted to warn you in your own interests".[29]

---

[23] "Sie sind ein gescheiter Junge, Einstein, ein ganz gescheiter Jung, Aber sie haben einen großen Fehler: sie Lassen sich nichts sagen!" Hoffmann Banesh and Dukas, Helen, *Albert Einstein Creator & Rebel*, 1973, New York: A Plume Book, p. 32.
[24] Seelig, 1956, p. 30; Seelig, 1954, p. 35.
[25] Seelig, 1956, p. 30; Seelig, 1954, p. 35.
[26] Seelig, 1956, p. 30; Seelig, 1954, pp. 35-36.
[27] *CPAE*, Vol. 1, p. 47.

Margarette von Uexküll reported in 1956 that one day in the experimental course of Pernet, "She had spent the whole of a warm June afternoon wrestling with an experiment in the Polytechnic's laboratory. Frustration overwhelming her, she was drawn into an argument with a small, fat physics Professor [Pernet], who refused to let her seal a test-tube with a cork for fear it would break. Suddenly she noticed 'a pair of the most extraordinary large shining eyes that were clearly warning me'. These belonged to Einstein, who quietly assured her that the professor was mad and had recently fainted during an angry fit in front of his class. He suggested that she give him her laboratory notes so that he could cook up some better results. At the next review, the professor exclaimed. 'There, you see. With a little goodwill, and despite my impossible methods, you can apparently work out something useful' ".[30]

Highfield and Carter mention, "It has to be said that von Uexküll may have varnished this anecdote, since she claimed that when Einstein borrowed her notebook he already had eight others awaiting similar doctoring".[31]

**2.2 Einstein never shows up or skips the classes of mathematicians**

Professor Adolf Hurwitz was a teacher and later Einstein's colleague at the ETH. Einstein took Hurwitz's two courses: Differential and Integral Calculus, and Integral Calculus.[32] Einstein wrote Marić in 1898 on Hurwitz's courses, "Hurwitz lectured on differential equations, except for partials, as well as Fourier series, and some on the calculus of variations and double integrals".[33]

Seelig writes that during the first three semesters in the Polytechnic "Einstein, together with Marcel Grossmann, attended the courses for engineers and mathematicians given with exemplary clarity by the German theoretician of numbers

---

and functions, Adolf Hurwitz. In personal relationships Hurwitz gave the impression of being rather reserved".[34]

Hurwitz was summoned to the Polytechnic from Königsberg in 1892 where he had worked as a 24-years-old youth as an associate professor. Seelig says that in the Polytechnic he initiated his pupils with exemplary clarity and precision into the mysteries of the differential calculus and the fundamentals of the theory of functions. Although for many years he suffered from painful kidney disease, he went on with his teaching even to the heroic extent of holding his seminar finally in his own home, enriching many fields of pure mathematics by his work. He had a great sense of humor and a love of music which he considered to be a complement to mathematics. He played the piano and chamber music was often played in his home. Later in 1908, when Einstein was to return to Zürich after his post in Bern in the Patent Office, he would join these musical sessions "with the whole chicken run!" (his wife and his two sons).[35]

Einstein had never shown up at the mathematical seminars of Hurwitz.[36] Hurwitz was a "talented mathematician", says Reiser. Indeed Einstein could have gotten good mathematical training from Hurwitz. "At that time", says Reiser, Einstein was less interested in mathematical speculation than in the visible process of physics".[37] Einstein felt that "the most fascinating subject at the time that I was a student was Maxwell's theory".[38] He found it difficult to accept for a long time the importance of abstract mathematics, and found high mathematics necessary only when developing his gravitation theory – he discovered the qualities of high mathematics around 1912.[39]

Einstein probably was not even very enthusiastic about Carl Friedrich Geiser's lectures,[40] which he skipped as much as he skipped Hurwitz's classes. Seelig spoke about "the outstanding Karl Friedrich Geiser from Bern who already had earned his title of professor at the age of 26. […] Geiser took great pains to make his lectures almost artistic".[41] Later, after he advanced the general theory of relativity, Einstein seemed to have been able to appreciate the value of Geiser's lectures.

---

[34] Seelig, 1956, pp. 28-29; Seelig, 1954, pp. 33-34.
[35] Seelig, 1956, p. 112; Seelig, 1954, pp. 132-133.
[36] Einstein to Adolf Hurwitz, September, 26, 1900, *CPAE*, Vol. 1, Doc 78. "Da ich mich wegen Mangels an Zeit nicht an dem mathematischen Seminar beteiligen konnte […]"
[37] Reiser, 1930, pp. 48-49.
[38] Einstein, Albert ,"Autobiographical notes" In Schilpp, Paul Arthur (ed.), *Albert Einstein: Philosopher-Scientist*, 1949, La Salle, IL: Open Court, pp. 1–95; p. 31.
[39] A girl named Barbara from Washington D.C. wrote Einstein on January 3, 1943, "I'm a little below average in mathematics", and she told Einstein that she had to work at it harder than most of her friends. Replying in English from Princeton on January 7, 1943, with his usual sense of humor, Einstein wrote in part as follows: "Do not worry about your difficulties in mathematics; I can assure you that mine are still greater". Dukas and Hoffmann, 1979, p. 8; Calaprice, Alice (ed), *Dear Professor Einstein, Albert Einstein's Letters to And From Children*, 2002, New York: Prometheus Books, pp. 139-140.
[40] Einstein took the following courses of Geiser: Analytical Geometry, Determinants, Infinitesimal Geometry, Geometry theory of Invariants, Exterior Ballistics. *CPAE*, Vol. 1, Doc. 28.

Indeed Einstein remembered the mathematician Carl Friedrich Geiser and his lectures of infinitesimal geometry in the second year in 1922: "I happened to remember the lecture on geometry in my student years [in Zürich] by Carl Friedrich Geiser who discussed the Gauss theory of surfaces. I found that the foundations of geometry had deep physical meaning in this problem".[42] The *elder* Einstein recalled, "I was fascinated by Professor Geiser's lectures on infinitesimal geometry, the true masterpieces of pedagogic art;[43] helped me later in the very struggle after the general relativity.[44]

During his studies at the Polytechnic Einstein also skipped another teacher's classes, the mathematician Herman Minkowski, who ten years later developed the mathematical formalism for Einstein's relativity theory. According to Seelig, Minkowski's "lectures, at times badly prepared but full of creative power".[45]

Einstein attended the following lectures of Minkowski: Geometry of Numbers, Function Theory, Potential Theory, Elliptic Functions, Analytical Mechanics, Variational Calculus, Algebra, Partial differential Equations, and Applications of Analytical Mechanics.[46]

Einstein's friend and biographer, Philipp Frank writes, "The teaching of mathematics was on a much higher level [than Weber's classes]. Among the instructors was Hermann Minkowski, a Russian by birth, who although still a young man, was already regarded as one of the most original mathematicians of his time. He was not very good lecturer, however, and Einstein was not much interested in his class".[47]

Minkowski, who was later the first to recognize the formal mathematical importance of Einstein's relativity theory, once admitted to his student, the physicist Max Born, "For me it came as a tremendous surprise, for in his student days Einstein had been a real lazybones. He never bothered about mathematics at all".[48] Einstein told his best friend Besso as late as 1916, "The study of Minkowski will not help you. His work is just extra complication".[49]

---

[41] Seelig, 1956, p. 29; Seelig, 1954, p. 34.
[42] Einstein, Albert, "How I Created the Theory of Relativity, translation to English by. Yoshimasha A. Ono, *Physics Today* 35, 1982, pp. 45-47; p. 47.
[43] Meisterstücke pädagogischer Kunst.
[44] Einstein, 1955, p. 11.
[45] Seelig, 1956, p. 27; Seelig, 1954, p. 32.
[46] *CPAE*, Vol. 1, Doc. 28.
[47] Frank, Philip, *Einstein: His Life and Times*, 1947, New York: Knopf, 2002, London: Jonathan, Cape, p. 20; Frank, Philip, *Albert Einstein sein Leben und seine Zeit*, 1949/1979, Braunschweig: F. Vieweg, p. 38.
[48] "Denn früher war Einstein ein richtiger Faulpelz. Um die Mathematik hat er sich überhaupt nicht gekümmert". Seelig, 1956, p. 28; Seelig, 1954, p. 33.
[49] Einstein to Besso, January 3, 1916, Einstein, Albert and Besso, Michele, *Correspondence 1903-1955* translated by Pierre Speziali, 1971, Paris: Hermann, Letter 13.

Einstein explained in his *Autobiographical Notes*,[50]

"I had excellent teachers (for example, Hurwitz, Minkowski), so that I should have been able to obtain a mathematical training in depth. I worked most of the time in physical laboratory, however, fascinated by the direct contact with experience. The balance of the time I used, in the main, in order to study at home the works of Kirchhoff, Helmholtz, Hertz, etc."

He saw that it was split into numerous specialties – so he did not know what to do,[51]

"Consequently, I saw myself in the position of Buridan's ass, which was unable to decide upon any particular bundle of hay. Presumably this was because my intuition was not strong enough in the field of mathematics to differentiate clearly the fundamentally important, that which is really basic, from the rest of the more or less dispensable erudition. Also, my interest in the study of nature was no doubt stronger; and it was not clear to me as a young student that access to a more profound knowledge of the basic principles of physics depends on the most intricate mathematical methods. This dawned upon me only gradually after years of independent scientific work".

**2.3 Einstein's Friends: Grossmann, Besso, and Marić**

During his Polytechnic student years Einstein made a few friends, Michele Angelo Besso, Marcel Grossmann, Friedrich Adler, and Mileva Marić with whom he married in 1903.

**Marcel Grossmann**

Einstein did not attend all the lectures of his Professors; he rather skipped many lessons, and before going to the two major examinations he had to pass in the four-year course, he studied instead from the notebooks of his good friend Marcel Grossmann. In 1955, Einstein told the story in his *Skizze*,[52]

"In these student days, I developed a real friendship with a fellow student Marcel Großmann. I solemnly went with him once a week to Café 'Metropol' on the Limmat embankment and talked to him not only about our studies, but also about anything that might interest young people whose eyes are open. He was not a kind of Vagabond and Eigenbrödler [loner] like me. […] Besides this he had just those gifts in abundance, which I lacked: quick learner and ordered in every sense. He not only visited instead of us in all eligible courses, but he also wrote them so neatly that he had printed his notebooks very well. In preparation for the exams he lent me these notebooks, which meant for me a lifesaver; what would have happened with me without it, I would rather not write and speculate".

---

[50] Einstein, 1949, pp. 14-15.
[51] Einstein, 1949, pp. 14-15.
[52] Einstein, 1955, p. 11.

In his *Autobiographical notes* Einstein acknowledged his debt to Grossmann without mentioning his name:[53]

"There were altogether only two examinations; aside from these, one could just about do as one pleased. This was especially the case if one had a friend, as did I, who attended the lectures regularly and who worked over their content conscientiously. This gave one freedom in the choice of pursuits until a few months before the examination, a freedom I enjoyed to a great extent, and I have gladly taken into the bargain the resulting guilty conscience as by far the lesser evil."

**Michele Angelo Besso**

Besso was six years older than Einstein. Besso was from a Jewish family; he was born near Zürich, and eventually was brought up in Italy. Besso left for Zürich and enrolled, in October 1891, in the mechanics section of the Zürich Polytechnic. There he took courses from the same professors who later taught Einstein. After four years of brilliant study he obtained his diploma in mechanical engineering and, soon afterwards, a position in an electrical machinery factory in Winterthur, near Zürich. Michele Besso came frequently to Zürich to attend musical soirées – he played the violin, like Einstein, and there he first met Einstein.[54]

Toward the end of 1896 or the beginning of 1897, during Einstein's first semester in the Polytechnic, he had met Besso at the Zürich home of a woman named Selina Caprotti, where people would meet to make music on Saturday afternoons. Besso possessed wide knowledge in physics, mathematics and philosophy, and he discussed with Einstein the philosophical foundations of physics.[55]

Later Besso worked with Einstein in the Patent Office in Bern, and after Einstein had left the Office for his first academic position, Besso continued to work there. In 1926 Einstein's friend in Zürich Heinrich Zangger heard that Besso's position in the Patent Office was in jeopardy. On December 12, 1926, from Berlin, Einstein wrote a few words in favor of his best friend.[56]

---

[53] Einstein, 1949, pp. 16-17.
[54] Speziali, Pierre, "Einstein writes to his best friend", in French, A. P. (ed), *Einstein A Centenary Volume*, 1979, London: Heinemann for the International Commission on Physics Education, pp. 263-269; pp. 263-264.
[55] Einstein to Besso, 6 March, 1952, in Einstein and Besso (Speziali), 1971, pp. 464-465.
[56] "Besso Stärke ist eine außergewöhnliche Intelligenz und unbeschränkte Hingabe an die berufliche und moralische Pflicht, seine Schwäche eine allzu geringe Entschlußfähigkeit. So erklärt es sich, daß sein äußerer Erfolg im Leben in keinem Verhältnis stand zu seinen glänzenden Fähigkeiten und seinem außergewöhnlichen Wissen auf technischem und rein wissenschaftlichem Gebiete. So erklärt es sich auch, daß im Amte zu wenig Aktenstücke seinen Namen tragen. Jeder im Amte weiß, daß er sich in schweirigen Fällen Rat bei Besso holen Kann; er versteht ungemein rasch die technische und juristische Seite jeder Patentangelegenheit und verhilft dem Kollegen gern zu einer raschen Erledigung,

Einstein wrote that, Besso's strength is an exceptional intelligence and unlimited devotion to the professional and moral duties. His weakness is low ability to make decisions. This explains why his success in life was disproportionate to his brilliant abilities and his extraordinary knowledge in purely and technical scientific fields. This also explains why few official documents bear his name. Everybody in the office who requested advices in engineering matters consulted him. He understands very quickly the technical and legal aspect of every patent case.

In fact, Besso was irreplaceable for Einstein, and the above traits were necessary for Einstein when he communicated with Besso and explained to him his new ideas.

Einstein goes on and says that Besso is most valuable as a consultant, and his withdrawal from the office would be a grave mistake. And Einstein tried to save his friend's position.[57]

And here some twenty years earlier Einstein described to Marić his schlemiel friend,[58]

"It's true, Michele is an awful schlemiel" and "The evening before last, Michele's director, with whom we are well acquainted, came over to play some music. He told us how completely useless and almost unbalanced Michele is, despite his extraordinary vast knowledge" Michele is so confused that "His manager sends him to the Castle power station to inspect and test the newly installed lines. Our hero decides to leave in the evening, to save valuable time of course, but he unfortunately misses the train. The next day he remembers his assignment too late. On the third day he gets to the train station on time, but to his horror realizes that he has forgotten what he is supposed to do. He immediately writes a card to the office saying that instructions should be wired! I don't think this fellow is normal".

**Freidrich Adler**

At this period the polytechnic enjoyed a great international reputation and had a large number of students from foreign countries. Among them were many eastern and southern European students who could not or would not study in their native countries. One of them was Freidrich Adler from Austria, who came to study in Zürich. He was a thin, pale, and blond young student. He had a fanatical faith in the revolutionary development of society. He was the son of Viktor Adler, the founder of

---

indem er sozusagen die Einsicht, der Andere den Willen bezw. Die Entschlußkraft zu dem Geschäft liefert. Hat er aber selbst eine Sache zu erledigen, dann ist der Mangel an Entschlußkraft hemmend. So entsteht die tragische Situation, daß einer der wertvollsten Arbeiter des Amtes, den ich in mancher Beziehung als nahezu unersetzlich bezeichnen möchte, nach außen den Eindruck mangelnder Brauchbarkeit machen muß".
Flückinger, 1960/1974, pp. 151-152.

[57] "Mein Ansicht ist also die, daß Besso eine höchst wertvolle konsultative Thätigkeit entfaltet und daß seine Entfernung aus dem Amte ein schwerer Fehler wäre. Ja noch mehr, sene technisch-juristischen Kenntnisse und sein Urteilsvermögen sind so außergewöhnlich, daß man es im Interesse des Staates nur bedauern kann, daß diese Fähigkeiten nicht an wichtigerer Stelle Verwendung finden als bei der formalen Bereinigung von Patentschriften". Flückinger, 1960/1974, pp. 151-152.

[58] Einstein to Marić, March 27, 1901, *CPAE*, Vol. 1, Doc. 94; Renn and Schulmann, 1992, letter 25.

the Social Democrat party of Vienna, who tried to keep his son out of politics by sending him to study physics in Zürich.[59] After graduation Einstein would meet Adler in 1908.

**Mileva Marić**

Mileva Marić was a Serbian from Novi Sad then in Hungary, who was three-and-a-half years older than Einstein. She was the only woman in the mathematical section of the school for Mathematics and Physics Teachers in the Polytechnic. Mileva Marić was Einstein's girlfriend and lover beginning 1899 and became his first wife on January 1903.

Hoffman and Dukas write that "At the Zürich Polytechnic, Einstein could not easily bring himself to study what did not interest him. Most of his time he spent on his own in joyful exploration of the wonderful of science, performing experiments and studying at first hand the works of great pioneers in science and philosophy. Some of these works he read with his Serbian classmate Mileva Marić, whom he later married."[60]

Einstein wrote Marić and told her about his readings during his student years. Classics of theoretical physics that Einstein studied were the works of Boltzmann ("the Boltzmann is absolutely magnificent", Einstein to Marić, 13 September 1900),[61] Helmholtz, Kirchhoff's "famous investigations of the motion of the rigid body" (Einstein to Marić, 1 August 1900),[62] and Hertz. "I returned the Helmholtz volume[63] and am now rereading Hertz's "Propagation of Electric Force"[64] with great care because I don't understand Helmholtz's treatise on the principle of least action in electrodynamics" (Einstein to Marić, 10 August 1899);[65] "but within a week I can have the municipal library send books by Helmholtz, Boltzmann, and Mach to me in Milan" (Einstein to Marić, 10, September, 1899).[66] During summer 1899 Einstein wrote Marić, "When I read Helmholz for the first time I could not – and still cannot – believe that I was doing so without you sitting next to me. I enjoy working together very much, and find it soothing and less boring".[67]

Milva Marić went to Zürich, because it was the second place in Europe that was willing to accept women to University. The University of Zürich and the Polytechnic were the first to accept German female students: Women from German speaking families went to Zurich and women from French speaking families like Marie Curie went to Paris. Almost twenty-one years old, she first studied a semester medical studies at the University of Zürich, and then in 1896, transferred to the Zürich Polytechnic, to the department which trained teachers of mathematics and physics.

---

[59] Frank, 1947/2002, p. 20; Frank, 1949/1979, p. 39.
[60] Hoffmann and Dukas, 1973, p. 28.
[61] Einstein to Marić, 13 Sept, 1900, *CPAE*, Vol. 1, Doc. 75; Renn and Schulmann, 1992, letter 21.
[62] Einstein to Marić, 1 Aug, 1900, *CPAE*, Vol. 1, Doc. 69; Renn and Schulmann, 1992, letter 15.
[63] Helmholtz, *Wissenschaftliche Abhandlungen*, Berlin, 1895.
[64] Hertz, *Untersuchungen Über Die Ausbreitung Der Elektrischen Kraft*, Leipzig,1892.
[65] Einstein to Marić, 10 Aug, 1899, *CPAE*, Vol. 1, Doc. 52; Renn and Schulmann, 1992, letter 8.
[66] Einstein to Marić, 10 sept, 1899, *CPAE*, Vol. 1, Doc. 54; Renn and Schulmann, 1992, letter 10; Presumably, Helmholtz, *Vorlesungen über die Elektromagnetische Theorie des Lichts*, Hamburg and Leipzig, 1897.
[67] Einstein to Marić, August 1899, *CPAE*, Vol. 1, Doc. 50, pp. 220-221; Renn and Schulmann, 1992, letter 7, p. 9.

Mileva met in Zürich a friend by the name Helene Savić, a history student from Vienna. Mileva and Helene lived in the same pension, of Fräulin Engelbrecht at Plattenstrasse 50, along with four other women, two Serbs and two Croats. Their friendship developed quickly. Besides living and studying together the women frequently joined to attend concerts and the theater, make excursions into the countryside around Zürich, and play their instruments. Sometimes the pension's lively atmosphere irritated some of the other residents. Milana Bota, a psychology student from Serbia, who was friendly with the women, complained to her parents that the noise often bothered her. [68]

One of the pension's most frequent visitors was Albert Einstein, who often joined in the women's musical performances. Einstein wrote Marić on October 10, 1899, about his visit to the pension: "I'm bringing a few luscious goodies from Mama, who promised to send us something for the household every so often: directly to Plattenstrasse 50. Pick up a copy of Helmholtz's electromagnetic theory of light in the meantime!". [69]

Einstein came to the pension with his violin and physics books. Mileva played the tamburitza (a mandolin-like-instrument), and later the piano, Helene, the piano, and Einstein, the violin. Milana Bota informed her parents:"Miss Marić has introduced me to her good friend, a German; his name is Einstein. He plays violin beautifully, he is a real artist, and so I'll have someone to play with again" (21 May 1898). Two weeks later she wrote them, "I wanted to answer your letter yesterday afternoon, but I had a visit, it was Miss Marić with the German I told you about and we made music the whole afternoon (3 June 1898)[70]

Mileva wore orthopedic boots to correct a physical deformity. She had been born with a dislocated left hip, leaving one leg shorter than the other. In a letter to her parents, Milana Bota described Mileva as "a very good girl, clever and serious, she is small, frail, dark, ugly, talks like a real Novi Sad girl, limps a little bit, but has very nice manners (18 March 1898).[71] Susanne Markwalder, the daughter of Frau Markwalder, in whose house Einstein lived, told Seelig, "At that time I also made the acquaintance of Frl. Mileva Marie who was in the finalists' class of the teachers' academy beside the Grossmünster. Einstein often worked with her. She was a modest, unassuming creature. Referring one day to her limb, one of Einstein's colleagues said: 'I should never have the courage to marry a woman unless she was absolutely sound'. Whereupon Einstein replied quite calmly: 'But she has such a lovely voice'!"[72]

---

[68] Popović, Milan, *In Albert's Shadow: The Life and Letters of Mileva Maric, Einstein's First Wife*, 2003, Maryland: The Johns Hopkins University Press, p. 3.

[69] Einstein to Marić, October 10, 1899, *CPAE*, Vol. 1, Doc. 63; Renn and Schulmann, letter 12. When Moszkowski later met Einstein in Berlin in 1916, he wrote, "Whenever Einstein talks of Helmholtz he begins in warm terms of appreciation, which tends to become cooler in the course of the conversation" [Wenn Einstein über Helmholtz spricht, so beginnt er mit einem glänzenden Auftakt, dessen Stimmung er im weiteren Verlauf nicht durchweg festhalten will] Moszkowski, Alexander, *Einstein the Searcher His Works Explained from Dialogues with Einstein*, 1921, translated by Henry L. Brose, London: Methuen & Go. LTD; appeared in 1970 as: *Conversations with Einstein*, London: Sidgwick & Jackson, 1970, p. 53; Moszkowski, Alexander, *Einstein, Einblicke in seine Gedankenwelt. Gemeinverständliche Betrachtungen über die Relativitätstheorie und ein neues Weltsystem. Entwickelt aus Gesprächen mit Einstein*, 1921, Hamburg: Hoffmann und Campe/ Berlin: F. Fontane & Co, p. 64.

[70] Popović, 2003, pp. 3-4.

Mileva Marić in fact never graduated from the Polytechnic. She failed the final examinations in "Funktionenth" and "Astronomie" due to poor grades in mathematics. Einstein finished first in his class in the intermediate exams of October 1898; second after him was his note taker Marcel Grossmann. Einstein wrote about the exams in his *Autobiographical Notes*: "[…] one had to cram all this stuff into one's mind for the examinations, whether one liked it or not. This coercion had such a deterring effect [upon me] that, after I had passed the examination, I found the consideration of any scientific problems distasteful to me for an entire year."[73]

In the final exams Einstein would seem to have relied too much on Grossman's lecture notes, because he did not repeat his success in the intermediate examination. He was awarded the diploma. Einstein submitted the required final diploma essay to Weber, which dealt with heat conduction. Seelig wrote: "Weber complained that for his work on Heat and transference, a subject which did not interest him, Einstein had not used the regulation paper. The professor insisted that he should copy the whole thing out again."[74] Small wonder that Einstein the rebel wrote an essay that did not interest him on a non formal paper.

**2.4 Einstein's residence in Zürich**

When Einstein arrived at Zürich on October 29th, 1896, he lived at the student quarter of Zürich 7; first with Frau Kägi at 4 Unionstrasse. There he remained for two years when he moved to the boarding house of Frau Stephanie Markwalder in 87 Klosbachstrasse and then finally with Hägi family at 17 Dolderstrasse.[75]

The primary school teacher, Susanne Markwalder, in whose mother's house Einstein lived as a student had told Seelig that her mother ran the boarding house above the Cantonal school. "In the evening, however, we [guests and students from different nationalities] made music and Einstein was our violinist. For preference he played Mozart and I did my best to accompany him on the piano".[76]

---

[71] Popović, 2003, pp. 4-5.
[72] Seelig, 1956, p. 38; Seelig, 1954, pp. 44-45.
[73] Einstein, 1949, pp. 15-17.
[74] Seelig, 1956, p. 30; Seelig, 1954, p. 35.
[75] Seelig, 1956, p. 35; Seelig, 1954, pp. 40-41.
[76] Seelig, 1956, pp. 35-36; Seelig, 1954, pp. 41-42.

Susanne Markwalder recalls, "He never annoyed my mother except when he forgot the house door key which incidentally he was constantly doing. The door bell would ring at the most impossible hours in the night and she would be woken up by the cry 'It's Einstein. I've forgotten my key again'. His impulsive and upright nature, however, was so irresistible that she never took long to forgive him. When he returned from his holidays in Milan he used to stand rather uncouthly in the doorway and say 'Will you have me back again or are you going to chuck me out?'"[77]

Of course Markwalder's report to Seelig of what happened 50 years earlier when Einstein was still anonymous, not a world icon and myth figure, should be critically read, as the other reports in Seelig's book. And thus one *can guess* that, Susanne Markwalder's mother was sometimes annoyed by Einstein's "impulsive and upright nature", and especially when he forgot the key.

## 3 Einstein was seeking a position

### 3.1 The rebel graduate Einstein is rejected

Einstein desired to be a theoretical physicist. He wanted to become an assistant to a professor at the Polytechnic. Einstein's sister, Maja reported in her biography of her brother Albert that, the assistant was selected by the professor and he helped the professor in the course by guiding the students, and also by assisting in experiments in the laboratories. After obtaining the diploma, Albert Einstein was equally qualified to assist in both these aspects, but none of his professors remembered their promises.[78] It became evident that the same professors who had praised his scientific interest and talent so highly had no intention whatsoever of taking him on as an assistant.[79]

Maja appears to have reported Einstein's subjective disappointment, which he perhaps expressed to her after graduation: he very likely felt that professors had promised him assistantship after he would graduate, and then none of them remembered their promises. Hoffmann and Dukas explain this: "By his forthrightness and his distrust of authority" Einstein "had alienated his professors, among them Heinrich Fridrich Weber". When Einstein "sought university positions, he was rebuffed.[80]

Reiser repeats the words of Maja, and also adds his own words, "It was clear that Albert Einstein would first turn to his professors. They had promised him an assistantship at the Polytechnic Academy, had even declared that they might establish a new position for him. When he reminded them of the matter, and explained the importance of this question for his whole life, his professors drew back timidly. He saw at once that they had forsaken him. Something must have happened to turn things against him, someone must have slandered him […]".[81]

---

[77] Seelig, 1956, p. 38; Seelig, 1954, p. 44.
[78] Winteler-Einstein, 1924, pp. 18-19.
[79] Frank, 1947/2002, p. 21; Frank, 1949/1979, p. 41.
[80] Hoffmann and Dukas, 1972, pp. 31-32.

Actually, the professors of department VI needed several assistants because of the large number of engineering students; and, as only a few students enrolled in the less lucrative fields such as mathematics and physics, virtually any graduate after his final exams could, if he wanted, become an assistant for a few years. This, however, did not apply to the rebel graduate Einstein. With Professor Pernet and Professor Weber he had no prospects – both were of course not fond of him at this stage (Weber was fond of Einstein in 1898 but later became alienated). Weber preferred not to become engaged with Einstein, and he took instead two mechanical engineers as his assistants. Einstein thus turned his hopes to the mathematicians. Einstein wrote his mathematics teacher Professor Adolf Hurwitz on September 23, 1900, asking him whether there is any prospect of becoming his assistant. Hurwitz was probably very surprised, because Einstein rarely had shown up at his seminars.[82]

Einstein had to confess his omissions in a letter dated September 26, 1900 to Hurwitz: "My most grateful thanks for your amiable letter! I am delighted to learn that there is some chance of me obtaining the position. Since through lack of time I was unable to attend the mathematical seminar where there was no chance of practice in practical and theoretical physics, I have nothing in my favor except the fact that I attended most of the seminars which offered me the opportunity. I think I must mention that in my student years I was mainly occupied with analytical mechanics and theoretical physics".[83]

**3.2 Professor Weber is Behind Einstein's Difficulties**

Einstein wrote Marić on March 27, 1901, "I am absolutely convinced that Weber is the blame". He wrote her, "I am convinced that under these circumstances it doesn't make any sense to write to any more professors, because they'll surely turn to Weber for information about me at a certain point, and he'll just give me another bad recommendation".[84]

And to his close friend Marcel Grossmann Einstein complained a month later on April 14 1901, "Dear Marcel, for the past three weeks I have been here with my parents and am trying from here to find a job as an assistant at some university. And I would have long found one had Weber not constantly played against me. Nevertheless, I am leaving no stone unturned and refuse to lose my sense of humor. God created the donkey and gave him a thick skin". And Einstein added: "As regards science, I have got a few wonderful ideas in my head which have to be worked out in due course".[85]

---

[81] Reiser, 1930, pp. 59-60.
[82] Einstein to Adolf Hurwitz, 23 September 1900, *CPAE*, Vol. 1, Doc. 77.
[83] Seelig, 1956, p. 49; Seelig, 1954, p. 58.
[84] Einstein to Marić, March 27, 1901, *CPAE*, Vol. 1, Doc. 94; Renn and Schulmann, 1992, letter 25, pp. 38- 39.

As to Grossmann, he was given an assistant post under Professor Wilhelm Fiedler who taught Descriptive geometry and Projective Geometry.[86] Grossman succeeded Fiedler as a professor in the Polytechnic in the autumn of 1907.[87]

Einstein graduated an excellent institute in Europe in which he could not stay as an assistant to any professor. Einstein thus wrote pleading letters to physicists all around Europe. He wrote Marić on April 4, 1901: "I will have soon graced all the physicists from the North Sea to the southern tip of Italy with my offer!"[88]

Desperate Einstein wrote on March 19, 1901 to the great physical chemist at the University of Leipzig, Professor Wilhelm Ostwald, who later won the Nobel Prize: "Since I was inspired by your book on general chemistry to write the enclosed article [on capillarity][89], I am taking the liberty of sending you a copy. On this occasion I venture also to ask you whether perhaps you might have use for a mathematical physicist who is familiar with absolute measurements". He wrote Ostwald, "I am taking the liberty of making such a request only because I am without means and only such a position would give me the possibility of further education", but he did not get any answer.

As the days passed and the postman brought no response, Einstein was even more desperate. On 3 April 1901 Einstein followed up his letter with a postcard saying how important the decision of his paper would be for him and – perhaps as a pretext for writing the postcard – wandering "I am not sure whether I have included my address" in Milan in the earlier letter, which in fact Ostwald had received.[90]

Still there was no response. On 12 April Einstein tried elsewhere, writing a brief note to Professor Heike Kamerlingh-Onnes in Leiden, Netherlands, again enclosing a reprint of his paper on capillarity. Nothing came of this application.[91]

---

[85] Einstein to Grossmann, April 14, 1901, *CPAE*, Vol. 1, Doc. 100.
[86] Seelig, 1956, p. 24; Seelig, 1954, p. 35. Otto Wilhelm Fiedler was a geometer. Einstein took two of his courses in the polytechnic: Central Projection and Projective Geometry. Einstein wrote Marić about Fiedler, "Fiedler is lecturing on projective geometry. He's the same indelicate tough person he always was, and a little impenetrable at that, though he's always brilliant and profound. In short: a master but unfortunately a terrible pedant too". *CPAE*, Vol. 1, Doc 28; Einstein to Marić, February 16, 1898, *CPAE*, Vol. 1, Doc. 39; Renn and Schulmann, 1992, letter 2. It is reasonable to assume that Einstein skipped Fiedler's courses.
[87] Seelig, 1956, p. 109; Seelig, 1954, p. 129.
[88] Einstein to Marić, 4 April 1901, *CPAE*, Vol. 1, Doc. 96; Renn and Schulmann, 1992, letter 26, p. 42.
[89] Einstein, Albert, "Folgerungen aus den Capillaritätserscheinungen", *Annalen der Physik* 309, 1901, pp. 513–523 ("Consequences of the Observations of Capillarity Phenomena"). This is Einstein's first published paper; Einstein submitted the paper in December 1900 and it was published in March 1901; Einstein to Ostwald, March 19, 1901, *CPAE*, Vol. 1, Doc. 92.
[90] Einstein to Wilhelm Ostwald, March 19, 1901, *CPAE*, Vol. 1, Doc. 92; Einstein to Wilhelm Ostwald, April 3, 1901, *CPAE*, Vol. 1, Doc. 95; Hofmann and Dukas, 1973, pp. 32-33.
[91] Einstein to Heike Kamerlinge Onnes, April 12, 1901, *CPAE*, Vol. 1, Doc. 98.

His father Hermann – the unsuccessful merchant, in ill health and a stranger to academic community, wrote Professor Ostwald on 13 April 1901,[92]

"I beg you to excuse a father who dares to approach you, dear Professor, in the interests of his son. I wish to mention first that my son Albert Einstein is 22 years old, has studied for four years at the Zürich Polytechnic and last summer brilliantly passed his diploma examinations in mathematics and physics. Since then he has tried unsuccessfully to find a position as assistant, which would enable him to continue his education […]".

And then Hermann went on with,

"Dear Professor, my son honors and reveres you the most among all the great physicists of our time", and finally Hermann pleaded with Ostwald to at least read Albert's capillarity paper, and "write him a few lines of encouragement so that he may regain his joy in his life and his work". And maybe he could obtain a position as assistant for his son…

Hoffmann and Dukas say that it is unknown whether Professor Ostwald wrote Einstein as result of this letter. What is known is that Einstein did not receive an assistantship.

**3.3 Einstein Finds Temporary Positions and Grossmann Rescues him**

In the meantime, Einstein had become a citizen of the canton of Zürich. His citizenship dates from February 21$^{st}$, 1901.[93] His chances of finding a job in Switzerland were now greater than they had been as a rebellious German Jew. Einstein finally found a temporary job. Starting May 19, 1901, he became a substitute teacher for two months at a high school in Winterthur. He was teaching five to six hours in the morning, and spent the afternoon working either in the library or at home.[94]

He wrote in a letter to Grossmann sent from Winterthur (probably on September 9, 1901), that he was working on Bolzmann's kinetic gas theory and, "the investigation of the relative motion of matter with respect to the luminiferous ether, a considerably simpler method had occurred to me, which is based on customary interference experiments". Einstein promised Grossman that when they see each other, "I will tell you about it".[95]

After Winterthur, another temporary position came his way. Dr jakob Nüesch advertised in the Swiss Teachers journal for a "Privat-Lehrer" to prepare a young man

---

[92] Hermann Einstein to Wilhelm Ostwald, April 13, 1901, *CPAE*, Vol. 1, Doc. 99; Hofmann and Dukas, 1973, pp. 33-34.
[93] Seelig, 1956, p. 51; Seelig, 1954, p. 60.
[94] Pais, Abraham, *Subtle is the Lord. The Science and Life of Albert Einstein*, 1982, Oxford: Oxford University Press, p. 46.
[95] Einstein to Grossmann, probably on September 6, 1901, *CPAE*, Vol. 1, Doc. 122.

Louis Cahen, from his boys' Realschule at Schaffhausen for the engineering section of the ETH. Einstein was finally engaged on the recommendation of his friend from Schaffhausen, Conrad Habicht. He was appointed for one year, to begin in September 1901, at a private school in Schaffhausen.[96]

Einstein wrote on December 18, 1901: "Since September 15, 1901, I am a teacher at a private school in Schaffhausen. During the first two months of my activities at that school, I wrote my doctoral dissertation on a topic in the kinetic theory of gases. A month ago I handed in this thesis to the University of Zürich".[97] This work was not accepted as a thesis, however. This setback, writes Einstein's biographer and colleague Abraham Pais, was the last one in Einstein's career. It came about the time he left Schaffhausen for Bern.[98] Einstein later (after 1919) spoke of the rejection of his dissertation as "a comic example of academic obscurantism".[99]

On May 3rd, 1901 (when Einstein had given the position at Winterthur), he wrote a letter from Milan to Professor Alfred Stern, telling Stern: "I have been appointed to teach mathematics at the Winterthur Technical School from May 15th to July 15th while the resident Professor does his military service. […] However, I received an offer. There is also a chance that later I shall obtain a permanent position in the Swiss Patent Office".[100]

Finally rescue came from Einstein's Polytechnic classmate, Marcel Grossmann. Grossman could not offer Einstein the assistantship he would like, as he was still only an assistant himself. But early in 1901 he had spoken with his father, Jules Grossmann, about Einstein's troubles, and his father strongly recommended Einstein to his friend Friedrich Haller, the railroad engineer and Director of the " Swiss Federal Office for Intellectual Property", "Eidgenössischen Amtes für geistiges Eigentum" – the Patent Office, "Patentamt", in Bern. After the first federal decree on patents, it was opened on November 15th, 1888 at 3 Lorrainestrasse and transferred in 1901 to the corner of the Bollwerk-Speichergasse. Haller was in charge of the Swiss Patent Office and its original director until 1921.[101]

Grossman then wrote Einstein that there was likely to be an opening for an examiner at the Swiss Patent Office in Bern.[102] In the letter to Grossman from April 14, 1901 Einstein wrote: "Please give my kind regards to your father and thank him for having taken the trouble and also for the confidence that he has shown in me by his recommendation" [to Friedrich Haller].[103] Einstein then wrote Mileva Marić on

---

[96] *CPAE*, Vol. 1, notes 3-4, p. 316.
[97] Einstein to the Swiss Patent Office, December 18, 1901, *CPAE*, Vol. 1, Doc. 129.
[98] Pais, 1982, p. 46.
[99] Plesch, John, *János, The Story of a Doctor*, translated by Edward Fitzgerald, 1949, New York, A.A. WYN, INC (first published 1947), p. 219.
[100] Einstein to Stern, May 3, 1901, *CPAE*, Vol. 1, Doc. 104; Seelig, 1956, p. 48; Seelig, 1954, pp. 58-59.
[101] Hoffmann and Dukas, 1973, p. 34.
[102] Hoffmann and Dukas, 1973, p. 34; Einstein to Grossman, 14 April, 1901, *CPAE*, Vol 1, Doc 100.
[103] Einstein to Grossmann, April 14, 1901, *CPAE*, Vol. 1, Doc. 100.

August 15, 1901, "The evening before last I got a letter from Marcel in which he informed me that I'll probably be getting a permanent position at the Swiss Patent Office in Bern! Isn't this too much to ask for all at once? Just think what wonderful job this would be for me! I'd be overjoyed if something came of it. Just think how nice it is of the Grossmanns once again to have taken the trouble of helping me."[104]

Einstein left Schaffhausen for Bern. Actually Einstein already understood that he would probably accept the position in the Patent Office in Bern and he quarreled with Nüesch. Einstein wrote Habicht from Bern on February 4, 1902, that he left Nüesch in Schaffhausen "with a bang".[105]

In 1936 when Einstein got the news from Zürich about Grossmann's death, he wrote his widow: "I remember our student days. He, the irreproachable student, I myself, unorderly and a dreamer. He, on good term with the teachers and understanding everything, I a pariah, discontent and little loved. But we were good friends and our conversations over iced coffee in the Metropole every few weeks are among my happiest memories. Then the end of our studies – I was suddenly abandoned by everyone, standing at a loss on the threshold life. But he stood by me and thanks to him and his father I obtained a post later with Haller in the Patent Office. It was a kind of salvation and without it, although I probably should not have died, I should have been intellectually damaged […]".[106]

Hoffmann and Dukas write,[107] that at some point Haller called Einstein for an interview in Bern. The interview lasted two hours. Haller quickly revealed Einstein's lack of relevant technical qualifications in patent law. Haller asked Einstein whether he possessed knowledge of patents, and Einstein replied, "no, nothing". Although Einstein lacked the technical qualification for a technical patent examiner, and understood nothing about intellectual property, Haller began to realize that there was something about the young man that transcended technicalities. There are strong reasons to believe that it was Einstein's rare mastery of Maxwell's electromagnetic theory that ultimately prompted Haller to offer him a provisional job in the Patent Office. Since there was no immediate opening, and since the law required that all openings be advertised,[108] this meant delay.

---

[104] Einstein to Marić, 15 April 1901, *CPAE*, Vol. 1, Doc. 101, Renn and Schulmann, 1992, letter 28.
[105] "mit Knalleffekt". Einstein to Habicht, February 4, 1902, *CPAE*, Vol. 1, Doc 133.
[106] Seelig, 1956, p. 208; Seelig, 1954, pp. 245-246.
[107] Hoffmann and Dukas, 1973, p. 34. Einstein grew in a technical environment in Munich and Milan: his father and uncle's Jacob Einstein electrical firm. Although Einstein lacked the technical qualifications in patents, he very likely did not entirely lack technical qualifications, such as understanding of operation of dynamos, electrical generators, and electrical equipment. And this could be a reason for why Haller decided to employ Einstein. In addition Einstein's mastery of Maxwell's electromagnetic theory also was a factor. Stachel, John, *Einstein's Miraculous Year. Five Papers that Changed the Face of Physics*, 2005, Princeton and Oxford: Princeton University Press, Introduction.
[108] The vacancy officially advertised in the *Schweizerisches Bundesblatt* (11 December 1901) listed the qualifications for the Patent Office Post as follows: "Through academic education in technical mechanics, or special leaning towards physics, a mastery of German and knowledge of French, or

Perhaps Einstein received some assurances of a position at that time.[109] In any event, he resigned his job at Schaffhausen and settled in Bern in February 1902, before he had any appointment there. At first his means of support were a small allowance from his family.[110] Then while waiting he did some private tutoring. Maurice Solovine his first pupil recalled later, "To help support himself he had to take pupils, who were hard to find and who did not bring in much money. He said to me one day that an easier way of earning a living would be to play the violin in public places".[111] On March 14 he became 23, and only on 23 June 1902 did Einstein start work at the Swiss Patent Office.[112]

*I wish to thank Prof. John Stachel from the Center for Einstein Studies in Boston University for sitting with me for many hours discussing special relativity and its history. Almost every day, John came with notes on my draft manuscript, directed me to books in his Einstein collection, and gave me copies of his papers on Einstein, which I read with great interest.*

---

mastery of French and knowledge of German, and possibly knowledge of Italian", stated the salary of 3500-4500 francs, and final date of 28 December for application is given. *CPAE*, Vol. 1, p. 327, note 1.
[109] Einstein to Marić, 22? July 1901, *CPAE*, Vol. 1, Doc. 119; Renn and Schulmann, 1992, letter 40.
[110] Pais, 1982, p. 46.
[111] Einstein, Albert, *Lettres à Maurice Solovine*, 1956, Paris: Gauthier Villars; *Letters to Solovine, with an Introduction by Maurice Solovine* (With an Introduction by Maurice Solovine, 1987), 1993, New York: Carol Publishing Group, p. VII; Solovine, Maurice, "Excerpts from a memoir", in French, 1979, p. 10.
[112] Hoffmann and Dukas, 1973, p. 35.